\documentclass[proof]{pasj00}
\usepackage{subfigure}
\usepackage{color}
\usepackage{ulem}
\draft

\begin{document}
\SetRunningHead{R. Usui et al.}{Outburst of LS V+44 17 Observed by MAXI and RXTE}
\Received{}
\Accepted{}
\Published{}

\title{Outburst of LS V+44 17 Observed by MAXI and RXTE, and Discovery of a Dip Structure in the Pulse Profile}

%

%
\author{%
  Ryuichi \textsc{Usui},\altaffilmark{1}
  Mikio \textsc{Morii},\altaffilmark{1}
  Nobuyuki \textsc{Kawai},\altaffilmark{1}
  Takayuki \textsc{Yamamoto},\altaffilmark{2,3}  
  Tatehiro \textsc{Mihara},\altaffilmark{2}
  Mutsumi \textsc{Sugizaki},\altaffilmark{2}
  Masaru \textsc{Matsuoka},\altaffilmark{2,4}
Kazuo \textsc{Hiroi},\altaffilmark{5}
Masaki \textsc{Ishikawa},\altaffilmark{6}
Naoki \textsc{Isobe},\altaffilmark{5,7}
Masashi \textsc{Kimura},\altaffilmark{8}
Hiroki \textsc{Kitayama},\altaffilmark{8}
Mitsuhiro \textsc{Kohama},\altaffilmark{4}
Takanori \textsc{Matsumura},\altaffilmark{9}
Satoshi \textsc{Nakahira},\altaffilmark{2}
Motoki \textsc{Nakajima},\altaffilmark{10}
Hitoshi \textsc{Negoro},\altaffilmark{3}
Motoko \textsc{Serino},\altaffilmark{2}
Megumi \textsc{Shidatsu},\altaffilmark{5}
Tetsuya \textsc{Sootome},\altaffilmark{2,11}
Kousuke \textsc{Sugimori},\altaffilmark{1}
Fumitoshi \textsc{Suwa},\altaffilmark{3}
Takahiro \textsc{Toizumi},\altaffilmark{1}
Hiroshi \textsc{Tomida},\altaffilmark{4}
Yoko \textsc{Tsuboi},\altaffilmark{9}
Hiroshi \textsc{Tsunemi},\altaffilmark{8}
Yoshihiro \textsc{Ueda},\altaffilmark{5}
Shiro \textsc{Ueno},\altaffilmark{4}
Kazutaka \textsc{Yamaoka},\altaffilmark{12}
Kyohei \textsc{Yamazaki},\altaffilmark{9}
Atsumasa \textsc{Yoshida}\altaffilmark{12}
}
\altaffiltext{1}{Department of Physics, Tokyo Institute of Technology, 2-12-1 Ookayama, Meguro-ku, Tokyo 152-8551}
\email{usui@hp.phys.titech.ac.jp}
\altaffiltext{2}{MAXI team, Institute of Physical and Chemical Research (RIKEN), 2-1 Hirosawa, Wako, Saitama 351-0198}
\altaffiltext{3}{Department of Physics, Nihon University, 1-8-14 Kanda-Surugadai, Chiyoda-ku, Tokyo 101-8308}
\altaffiltext{4}{ISS Science Project Office, Institute of Space and Astronautical Science (ISAS), Japan Aerospace Exploration Agency (JAXA), 2-1-1 Sengen, Tsukuba, Ibaraki 305-8505}
\altaffiltext{5}{Department of Astronomy, Kyoto University, Oiwake-cho, Sakyo-ku, Kyoto 606-8502}
\altaffiltext{6}{School of Physical Science, Space and Astronautical Science, The graduate University for Advanced Studies (Sokendai), Yoshinodai 3-1-1, Chuo-ku, Sagamihara, Kanagawa 252-5210}
\altaffiltext{7}{Institute of Space and Astronautical Science (ISAS), Japan Aerospace Exploration Agency (JAXA) ,3-1-1 Yoshino-dai, Chuo-ku, Sagamihara, Kanagawa 252-5210}
\altaffiltext{8}{Department of Earth and Space Science, Osaka University, 1-1 Machikaneyama, Toyonaka, Osaka 560-0043}
\altaffiltext{9}{Department of Physics, Chuo University, 1-13-27 Kasuga, Bunkyo-ku, Tokyo 112-8551}
\altaffiltext{10}{School of Dentistry at Matsudo, Nihon University, 2-870-1 Sakaecho-nishi, Matsudo, Chiba 101-8308}
\altaffiltext{11}{Department of Electronic Information Systems, Shibaura
Institute of Technology, 307 Fukasaku, Minuma-ku, Saitama, Saitama 337-8570}
\altaffiltext{12}{Department of Physics and Mathematics, Aoyama Gakuin University,\\ 5-10-1 Fuchinobe, Chuo-ku, Sagamihara, Kanagawa 252-5258}
\KeyWords{stars: pulsars: individual (LS V +44 17, RX J0440.9+4431) --- stars: neutron --- X-rays: binaries} 

\maketitle

\begin{abstract}
 We report on the first observation of an X-ray outburst of a Be/X-ray
 binary pulsar LS V +44 17/RX J0440.9+4431, and the discovery of an
 absorption dip structure in the pulse profile. 
 An outburst of this source was discovered by MAXI GSC in 2010 April.
 It was the first detection of the transient activity of LS V +44 17 since the source
 was identified as a Be/X-ray binary in 1997. 
 From the data of the follow-up RXTE observation near the peak of the outburst, we
 found a narrow dip structure in its pulse profile which was clearer in the lower energy bands. 
 The pulse-phase-averaged energy spectra in the 3$-$100 keV band can be
 fitted with a continuum model containing a power-law function with an exponential cutoff and a blackbody component, which are modified at low energy by an absorption component.
 A weak iron K$\alpha$ emission line is also detected in the spectra.
From the pulse-phase-resolved spectroscopy we found that the absorption column density at the dip phase was much higher than those in the other phases.
 The dip was not seen in the subsequent RXTE observations at lower flux levels.
These results suggest that the dip in the pulse profile originates from the
 eclipse of the radiation from the neutron star by the accretion column.
\end{abstract}

\section{Introduction}

LS V +44 17/RX J0440.9+4431 is a Be/X-ray binary pulsar
which consists of a neutron star and a main-sequence Be star.
Be/X-ray binary shows periodic and/or non-periodic outbursts 
which are thought to be caused by a mass accretion from the Be star to the neutron star 
when the neutron star pass through the circumstellar disk around the companion Be star at the periastron \citep{okazaki2001}.

The binary system LS V +44 17 was discovered from the search for the positional cross-correlation between the SIMBAD OB star catalog and the ROSAT all-sky survey sources located at low galactic latitude ($|b| < 20^{\circ}$) \citep{motch1997}.
\citet{reig1999b} observed the X-ray source with RXTE and confirmed it as an X-ray binary pulsar with a pulse period of $202.5\pm 0.5$ s.
They also revealed that the pulse profile was sinusoidal and
the spectrum was represented by a power-law with a photon index = 2, an absorption column density of $N_{\mathrm{H}} = 4.1 \times 10^{22}\ \rm{cm}^{-2}$,  and an exponential higher-energy cutoff at $E_{\rm{fold}}$ = 1.9 keV. The cutoff energy is rather low compared with those of the typical Be/X-ray pulsars, 10$-$20 keV.
They argued that 
LS V +44 17 belongs to a class of persistent Be/X-ray binaries,
which are characterized by no outburst activity, the low luminosity ($L_x \leq 10^{34-35}\ \mathrm{ergs\ s^{-1}}$), and the long pulse period in a wide orbit.
These features are large contrasts to those of transient Be/X-ray binaries 
such as A 0535+262 and EXO 2030+375.

The optical companion star LS V +44 17 (VES 826, BSD 24$-$491) is classified as a B0.2Ve star \citep{belczynski2009}.
\citet{reig2005} performed optical and infrared observations throughout almost 10 years
and estimated the distance of LS V +44 17 to be $3.3 \pm 0.5$ kpc.
We use the source distance through this paper.
The long-term observation showed a double-peaked H$\alpha$ line whose equivalent width varied.

LS V +44 17 has no record of significant activity until the first detection of an outburst by MAXI in 2010 March \citep{morii2010}.
We report the results of the MAXI observation and the follow-up observations by RXTE in this paper.

\section{Observation and Data Reduction}

\subsection{Discovery of the Outburst by MAXI}

Monitor of All-sky X-ray Image (MAXI; \cite{matsu2009}) has been operational since 2009 August. 
MAXI carries two types of X-ray cameras: the Gas Slit Camera (GSC) and the Solid-state Slit Camera (SSC).
The GSC employs gas proportional counters with one-dimensional position sensitivity as X-ray detectors and covers the energy band of 2$-$20 keV \citep{mihara2011}.
On the other hand, the SSC employs the X-ray CCD arrays and covers
the energy band of 0.5$-$12 keV \citep{tomida2011}.
In this paper, we use only the GSC data because the SSC did not cover the field of LS V +44 17 during the outburst.

\begin{figure}[p]
  \begin{center}
    \FigureFile(160mm,160mm){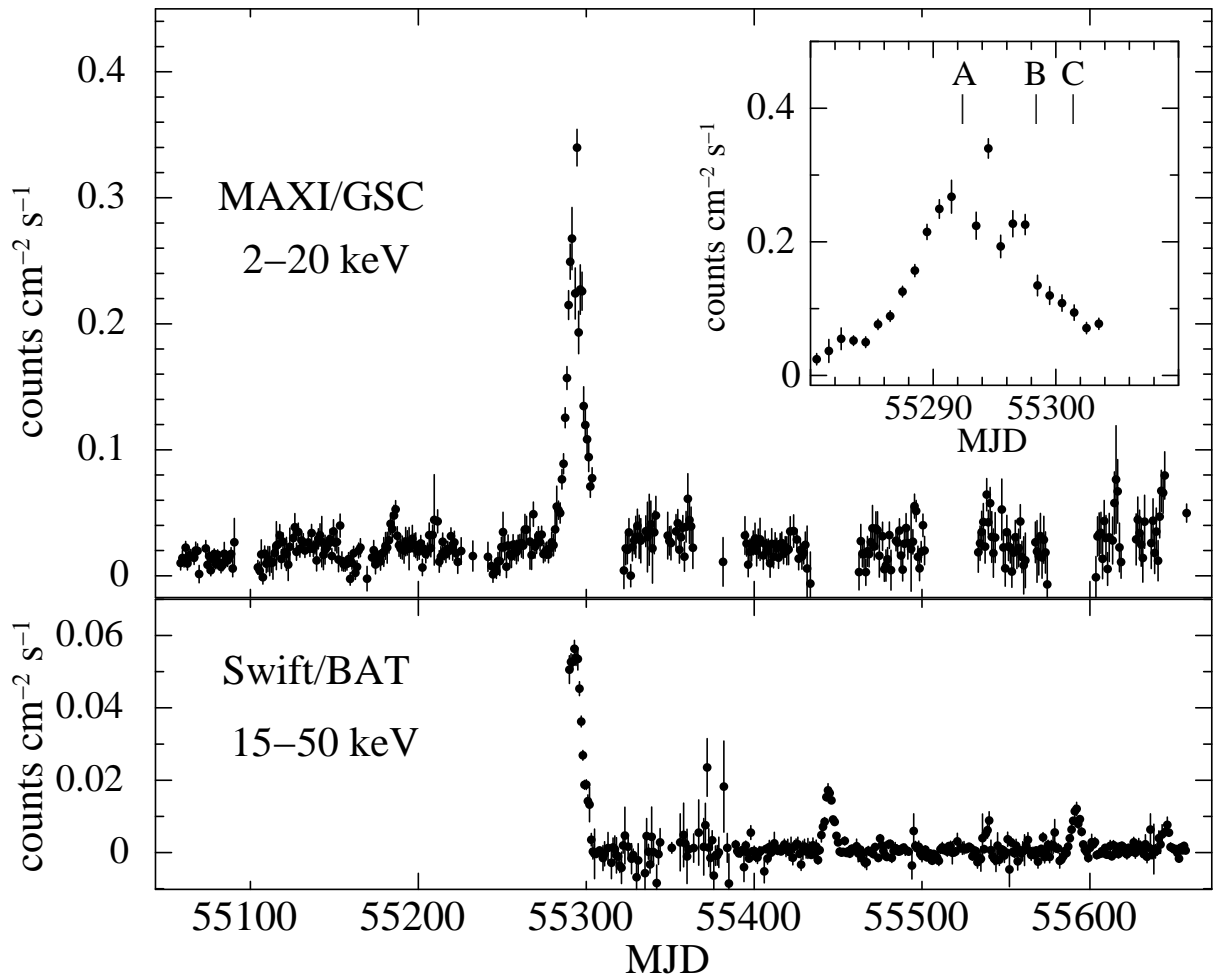}
  \end{center}
  \caption{X-ray Light curves of LS V +44 17. Top panel: MAXI/GSC light curve from 2009 August 15 to 2011 April 6. The outburst is zoomed in the inset, where the RXTE observation times are indicated with three vertical bars. Bottom panel: Swift/BAT light curve from 2010 April 4 to 2011 April 6. For reference, the flux of the Crab pulsar is 2.5 and 0.22 $\mathrm{counts}\ \mathrm{cm}^{-2}\ \mathrm{s}^{-1}$ in the top and bottom panels, respectively.}\label{fig:lsv_lc}
\end{figure}

Figure \ref{fig:lsv_lc} (top panel) shows a light curve of the source 
obtained by the GSC from 2009 August 15 to 2011 April 6.
The GSC data is obtained from the MAXI archival products distributed via a MAXI web site at RIKEN\footnote{http://maxi.riken.jp/},
where all the in-orbit calibration information described in \cite{sugizaki2011} is applied.
On 2010 March 31, MAXI detected an outbursting X-ray source whose position is consistent with that of LS V +44 17 \citep{morii2010}.
On April 1, 
Swift-BAT independently detected the outburst and automatically triggered the XRT follow-up observation, and then confirmed the source position to be consistent with LS V +44 17 \citep{stratta2010}.
The 205-s coherent pulsation was also detected in the XRT data (Kennea, private communication).
Since the first detection, X-ray flux had increased for a week and finally reached $\sim$150 mCrab in the 2-20 keV band of the GSC.

Figure \ref{fig:lsv_lc} (bottom panel) shows the BAT light curve\footnote{http://swift.gsfc.nasa.gov/docs/swift/results/transients/} from 2010 April 4 to 2011 April 6 in the 15$-$50 keV band.
The second and third outbursts are shown around MJD 55442 \citep{krivonos2010} and 55592 \citep{tsygankov2011}, respectively.
MAXI was not able to observe these outburst because the source had been in the unobservable area by the GSC for the periods. 

\subsection{RXTE Observation and Data Reduction}

RXTE performed a follow-up observation 
on 2010 April 6
and confirmed the 205 s pulsation \citep{finger2010}.
During the outburst,
two more observations were carried out on April 12 and 15.
Table \ref{tab:obs} summarizes logs of the three observations.
Hereafter,
these three RXTE observations are referred to as Obs.\,A, B and C, whose epochs are indicated by the ticks in figure \ref{fig:lsv_lc} (inset).
During Obs.\,A, 
the luminosity was near the peak level
($L_{\mathrm{X}} \sim 7.1 \times 10^{36}\ \mathrm{erg\ s^{-1}}$ in 3$-$30 keV)
while those in Obs.\,B and Obs.\,C decreased to $L_{\mathrm{X}} \sim 4.7 \times 10^{36}\ \mathrm{erg\ s^{-1}}$ and 
$\sim 2.2 \times 10^{36}\ \mathrm{erg\ s^{-1}}$, respectively.
Here we assume the source distance of 3.3 kpc.
These observations provided the data with high photon statistics  
in the 2$-$60 keV band with the Proportional Counter Array (PCA; \cite{jahoda2006}) and
in the 15$-$250 keV band with the High-Energy X-ray Timing Experiment (HEXTE; \cite{rothschild1998}). 
We then carried out timing and spectral analyses for all the observations 
and compared the pulse profiles and spectral parameters.

We obtained the archived data from NASA/HEASARC and reduced the data with the standard analysis methods,
using HEASOFT version 6.10 and 
CALDB (calibration database) version 20100309 provided by NASA/GSFC RXTE GOF.
The event data of the PCA and HEXTE were screened by the status of the satellite as follows: 
the pointing offset smaller than 0.02 degrees, 
the elevation angle of the satellite larger than 10 degrees and 
the time since passage of South Atlantic Anomaly more than 30 minutes.
We then applied the barycentric correction to the screened event data using \texttt{faxbary}.

The PCA light curves and spectra were extracted from the top layer of PCU2 alone to obtain the high signal-to-background data in the 3$-$30 keV energy band.
The background spectra were estimated with \texttt{pcabackest} 
using the background model for ``bright" sources which was released on 2005 November 28.
In the spectral analysis, 
we used the response matrices made by \texttt{pcarsp} for each observation.
We added systematic error of 0.5\% to the spectra, 
following the recommendation of the PCA team\footnote{http://www.universe.nasa.gov/xrays/programs/rxte/pca/doc/rmf/pcarmf-11.7/}.

The HEXTE source spectra in the 20$-$100 keV band were extracted from the Science Event data of Cluster-A, 
while the background spectra were estimated from the data of Cluster-B using \texttt{hextebackest}.
Since the HEXTE background spectra have a relatively large uncertainty\footnote{RXTE news http://heasarc.gsfc.nasa.gov/docs/xte/xhp\_new.html} 
around 63 keV, we ignored the energy range of 57$-$74 keV in the spectral analysis. 
The exposure time correction was applied for the source and the background spectra
considering the dead time of the instrument.
Response matrices used in the HEXTE spectral analysis was made with \texttt{hxtrsp}.
To reduce small concave residuals shown in the background-subtracted spectra, 
which are presumably caused by the unadjusted normalization of the model background spectrum,
we calibrated the scale factor of the background by fitting the spectrum with the power-law function
with \texttt{cornorm} in XSPEC.
We obtained the background scale factor to be 0.105 and fixed it in the following spectroscopy.

\section{Analysis and Results}

\subsection{Timing Analysis}

We estimated the pulse periods of LS V +44 17 from the 3$-$20 keV PCA light curves obtained in Obs.\,A, B and C.
Since any orbital elements of the binary system are not known, 
we did not apply any correction for the orbital motion.
We estimated the pulse periods with the method of the epoch folding search 
around the period of 202.5 s reported in Reig and Roche (\yearcite{reig1999b}). 
The obtained periods are $205.0 \pm 0.4$ s, $205.0 \pm 0.7$ s and $204.8 \pm 0.7$ s for the three observations, respectively.
The errors of the periods represent the 90 \% confidence limits, which are obtained with Monte Carlo simulations.

\begin{figure}[p]
  \begin{center}
    \FigureFile(160mm,160mm){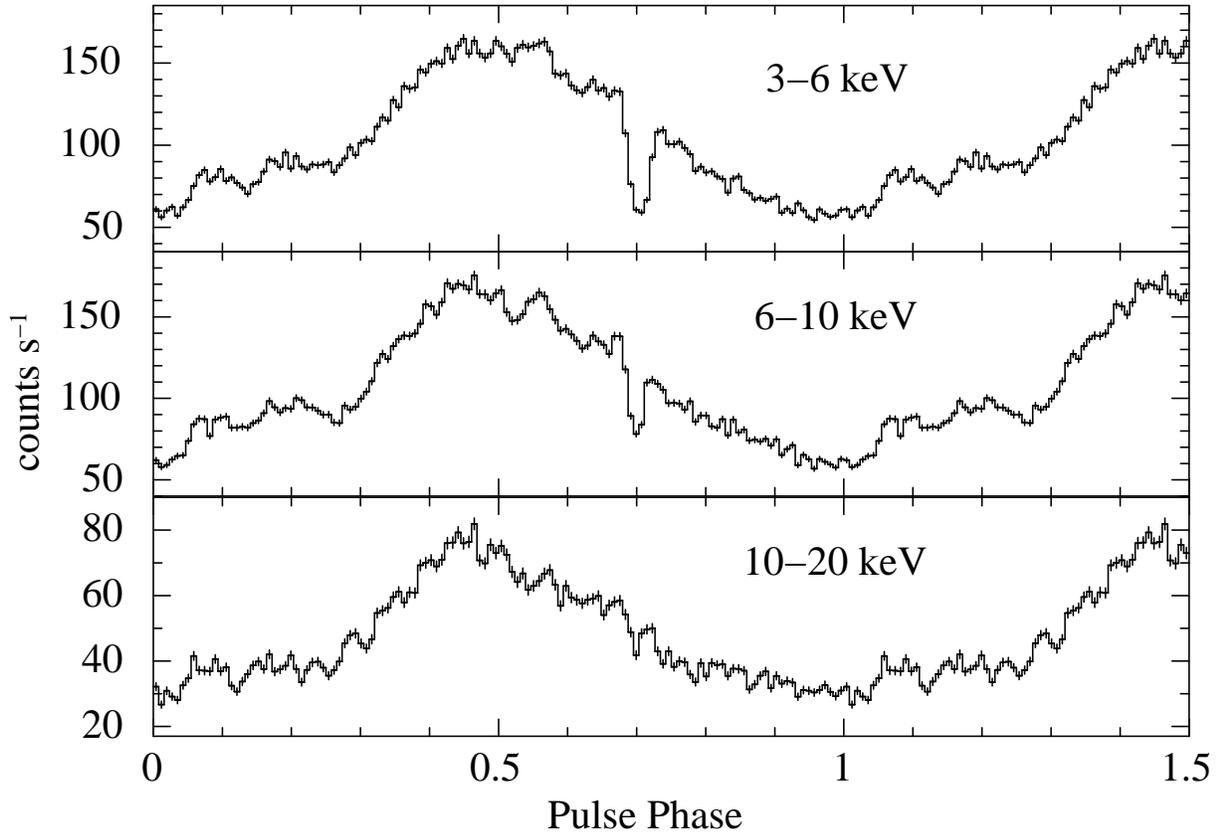}
  \end{center}
  \caption{PCA pulse profiles of LS V +44 17 at the 3$-$6, 6$-$10 and 10$-$20 keV band during Obs.\,A, where the number of bin is 128 bin and the error bars correspond to 1 sigma statistical error.
  	A half of the pulse cycle is repeated for clarity.
                  The sharp dip features are seen in the 3$-$6 and 6$-$10 keV energy bands at the phase $\sim 0.7$ (top and middle panels). }
  \label{fig:pulse1}
\end{figure}

\begin{figure}[p]
  \begin{center}
    \FigureFile(160mm,160mm){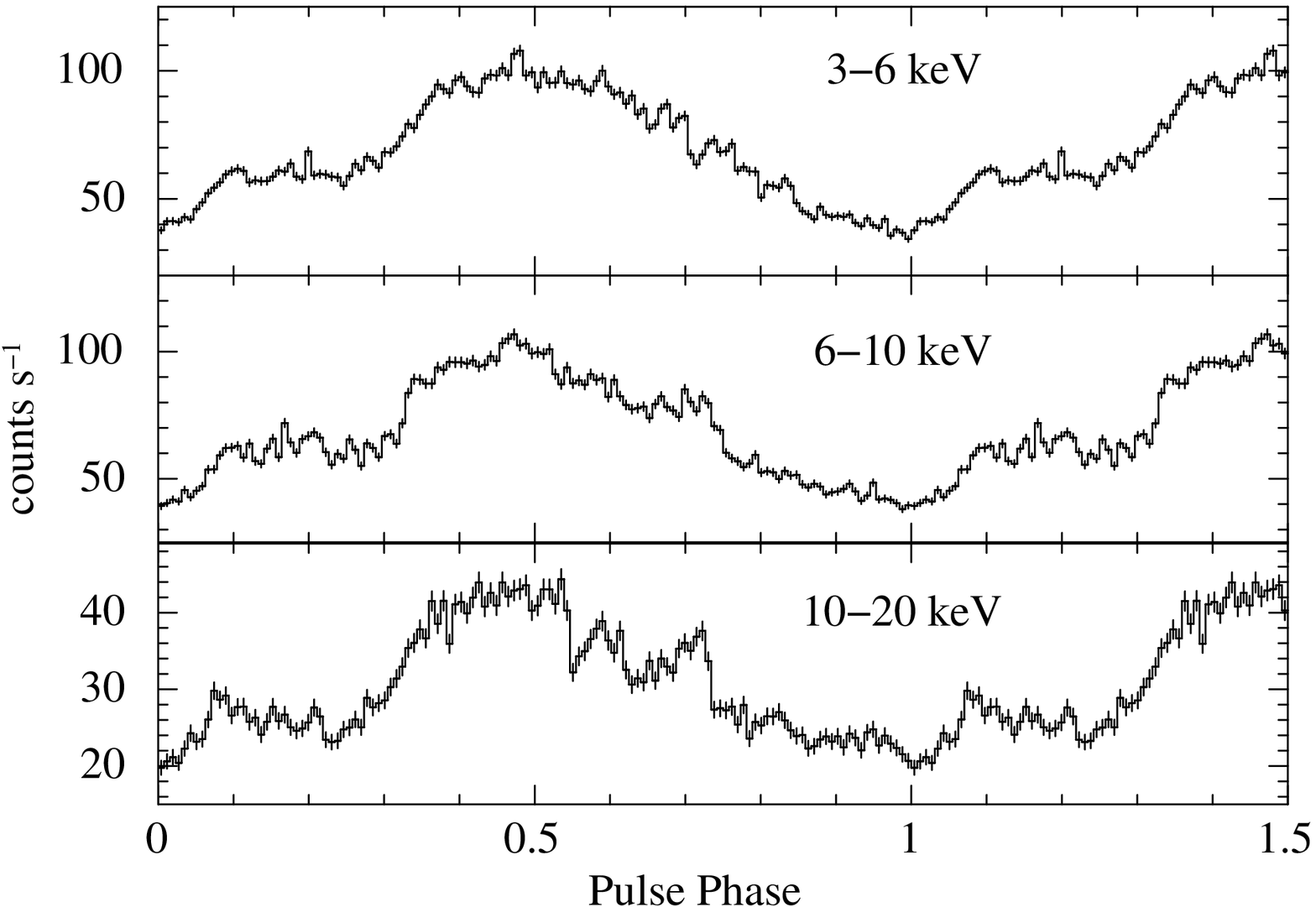}
  \end{center}
  \caption{Pulse profile of LS V +44 17 during Obs.\,B}\label{fig:pulse2}
\end{figure}

\begin{figure}[p]
  \begin{center}
    \FigureFile(160mm,160mm){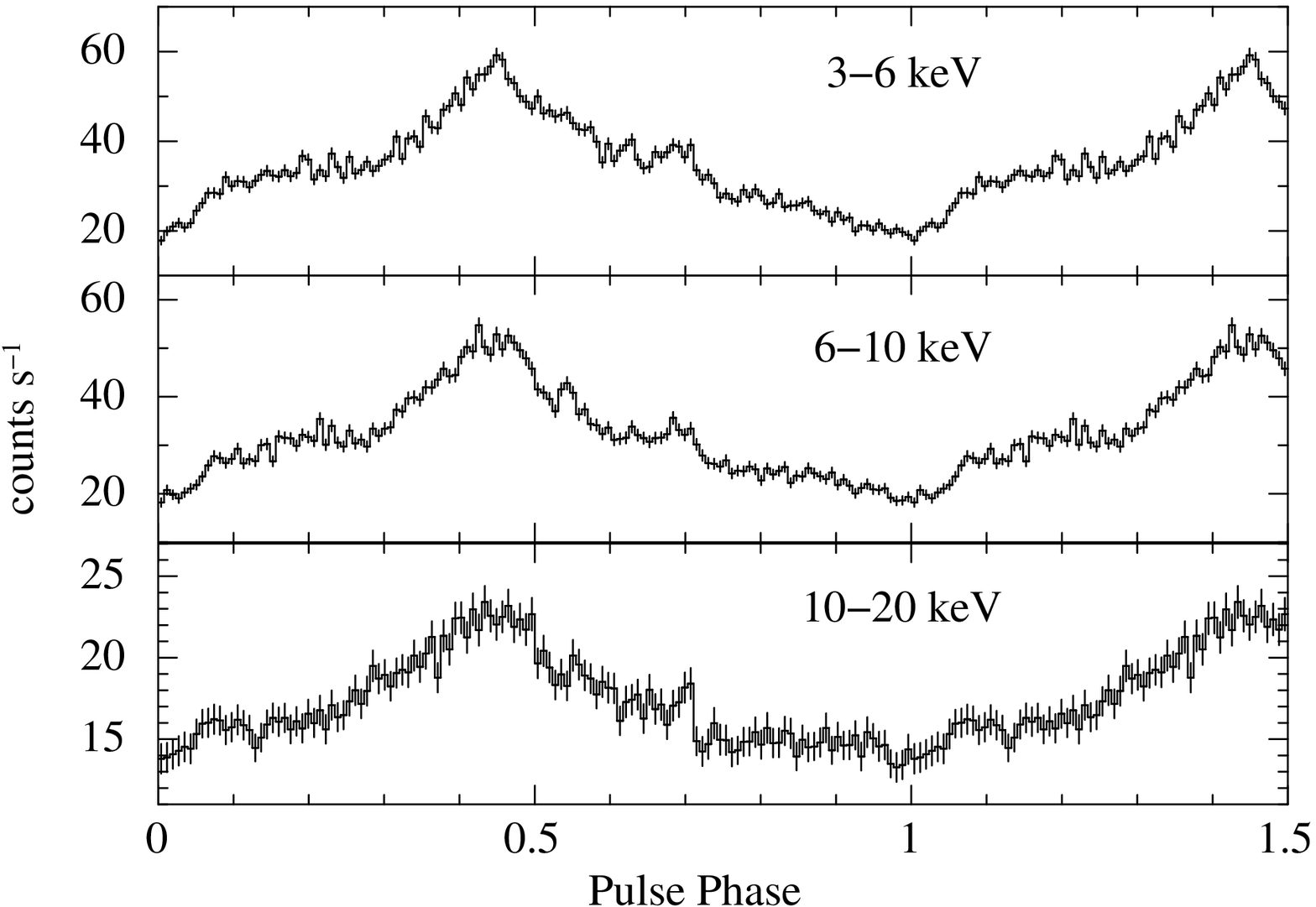}
  \end{center}
  \caption{Pulse profile of LS V +44 17 during Obs.\,C}\label{fig:pulse3}
\end{figure}

With the estimated period of 205 s,
we obtained the pulse profiles in the three energy bands, 3$-$6, 6$-$10 and 10$-$20 keV for the three PCA observations, which are shown in figures \ref{fig:pulse1}, \ref{fig:pulse2} and \ref{fig:pulse3}, where the background is not subtracted.
The phase zero was chosen at the minimum 
of the 3$-$20 keV pulse profiles for each observation. 
The shapes of the pulse profiles are similar to those observed in 1998 (see \cite{reig1999b}).
Notably a narrow dip structure is seen near the phase 0.7 
in the profile of Obs.\,A (figure \ref{fig:pulse1}).
The depth of the dip is shallower in the higher energy bands.

\subsection{Pulse-phase-averaged Spectroscopy}

\begin{figure}[p]
  \begin{center}
    \FigureFile(160mm,160mm){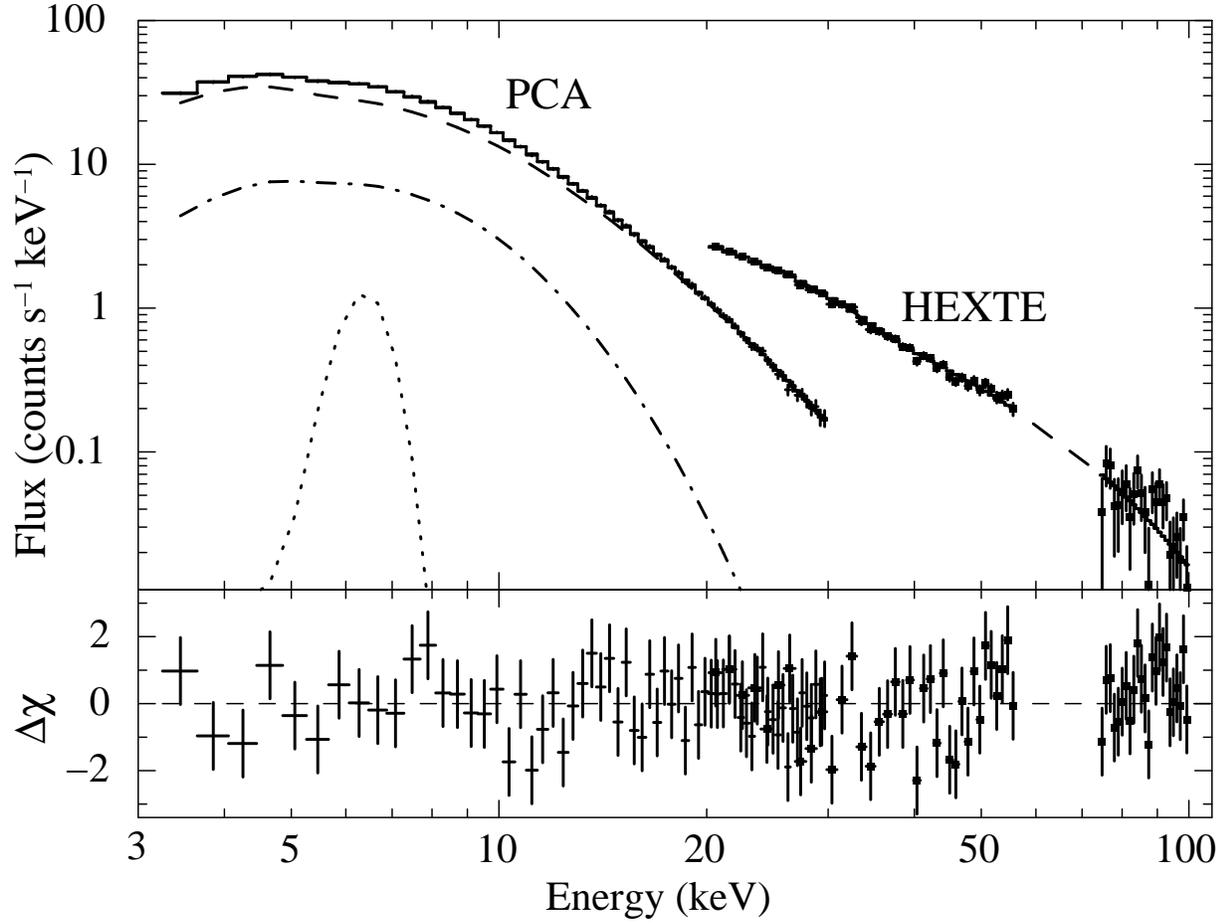}
  \end{center}
  \caption{The pulse-phase-averaged spectrum of LS V +44 17 observed by PCA (3$-$30 keV) and HEXTE (20$-$100 keV) in Obs.\,A. The solid histogram represents a model consisting of a power law component with exponential cut-off (PL: dash line), a blackbody component (BB: dash-dot line) and a gaussian function (Fe: dot line). The residuals are shown in the lower panel with a unit of one sigma.}\label{fig:spectrum}
\end{figure}

We analyzed pulse-phase-averaged spectra for each of the three RXTE observations 
using both of the PCA (3$-$30 keV) and the HEXTE (20$-$100 keV) data.
At first, 
we tried to fit the spectra with an absorbed power-law model modified with an exponential cut-off (PL), 
which is often used to fit the spectra of accreting pulsars as also in \citet{reig1999b}.
The fitting results, however, were not acceptable 
with the reduced chi-squared ($\chi^2_{\mathrm{red}}$) $>$  2.0, where large residuals remained in the 3$-$10 keV band. 
We obtained an acceptable fit by adding a blackbody component (BB) and a gaussian function (Fe) representing a neutral iron emission line at $E_{Fe} \sim$ 6.4 keV with the width fixed at 0.1 keV.\footnote{This width is much smaller than the energy resolution of the detector ($<$ 18\% at 6 keV).}
Throughout our paper, we refer this model as PL+BB+Fe .
Figure \ref{fig:spectrum} shows the pulse-phase-averaged spectrum during Obs.\,A, and 
table \ref{tab:spec_pars} summarizes the parameters obtained by the fits for the three observations.
The photon indices of the power-law ($\alpha$) showed no significant variation among the observations, while 
the folding energy ($E_{\mathrm{fold}}$) became slightly lower as the flux decreased.
The radius of the blackbody ($R_{\mathrm{BB}}$) did not change
significantly at $\sim$ 0.5$-$0.6 km,  while the temperature ($kT_{\mathrm{BB}}$) decreased.
Be/X-ray binaries typically have a weak iron emission line at $\sim$ 6.4 keV with an equivalent width $W_{eq} <$ 100 eV (e.g. \cite{Reynolds2010} for A 0535+262 and \cite{naik2011} for GRO J1008$-$57). Compared to the equivalent widths of other Be/X-ray binaries, the obtained value of LS V +44 17, $\sim 40 \rm{eV}$, is not unusual and consistent with that derived by \citet{reig1999b}.

\subsection{Pulse-phase-resolved Spectroscopy}

\begin{figure}[p]
  \begin{center}
    \FigureFile(160mm,160mm){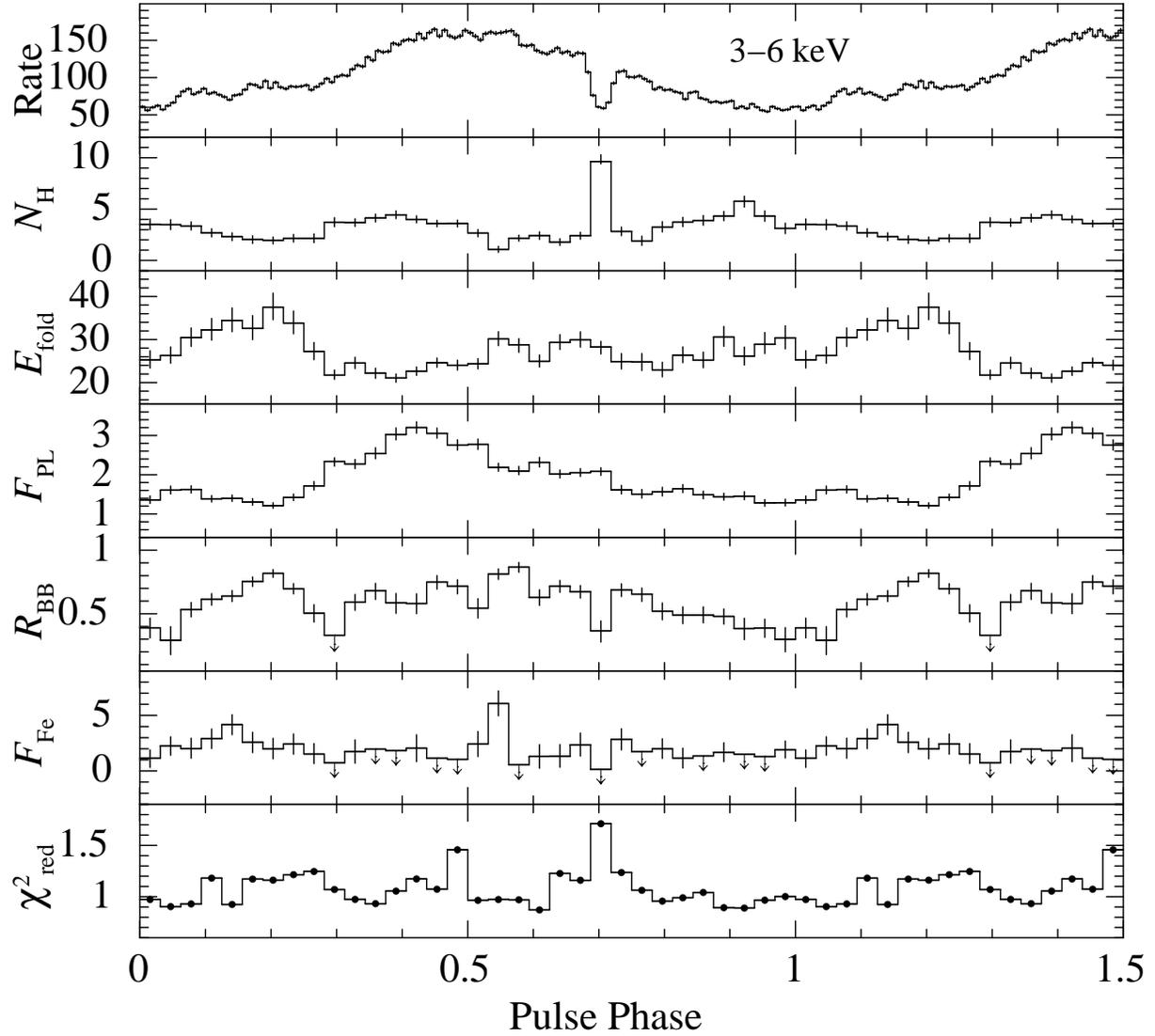}
  \end{center}
  \caption{Phase dependence of spectral parameters and $\chi^2_{\mathrm{red}}$ obtained from pulse-phase-resolved spectroscopy during Obs.\,A. The units of parameters are same as shown in table \ref{tab:spec_pars}. The upper panel shows the pulse profile in 3$-$6 keV. Errors are given at 1-sigma confidence limits. Downward arrows are the 1-sigma upper limits. $F_{\rm{PL}}$ and $F_{\rm{Fe}}$ are flux of the power-law and iron line component}. \label{fig:res_pars}
\end{figure}

To investigate the dip structure found in the pulse profile during Obs.\,A, 
we analyzed pulse-phase-resolved spectra during the period.
We divided the full pulse cycle into 32 phase intervals, 
where the phase 0.69$-$0.72 corresponds to the dip phase.
We used the same model that was applied for the pulse-phase-averaged spectra, PL+BB+Fe model, whose photon index, blackbody temperature and gaussian peak energy are fixed at the best-fit values as shown in table \ref{tab:spec_pars}, $\alpha = 1.15$, $kT_{\mathrm{BB}} = 2.17$ keV and $E_{\mathrm{Fe}} = 6.44$ keV. 
Figure \ref{fig:res_pars} shows the phase dependence of spectral parameters and the reduced chi-squared ($\chi^2_{\mathrm{red}}$).
The pulse profile of the 3$-$10 keV is also shown for a comparison.
The most remarkable result is the largest column density at the dip phase.
We note that the fit is not acceptable at the dip phase with a $\chi^2_{\mathrm{red}}$ of 1.66 for 111 degree of freedom (see table \ref{tab:dip_pars}).
Thus we examined the dip spectrum further in detail.

\begin{figure}[p]
  \begin{center}
    \FigureFile(160mm,160mm){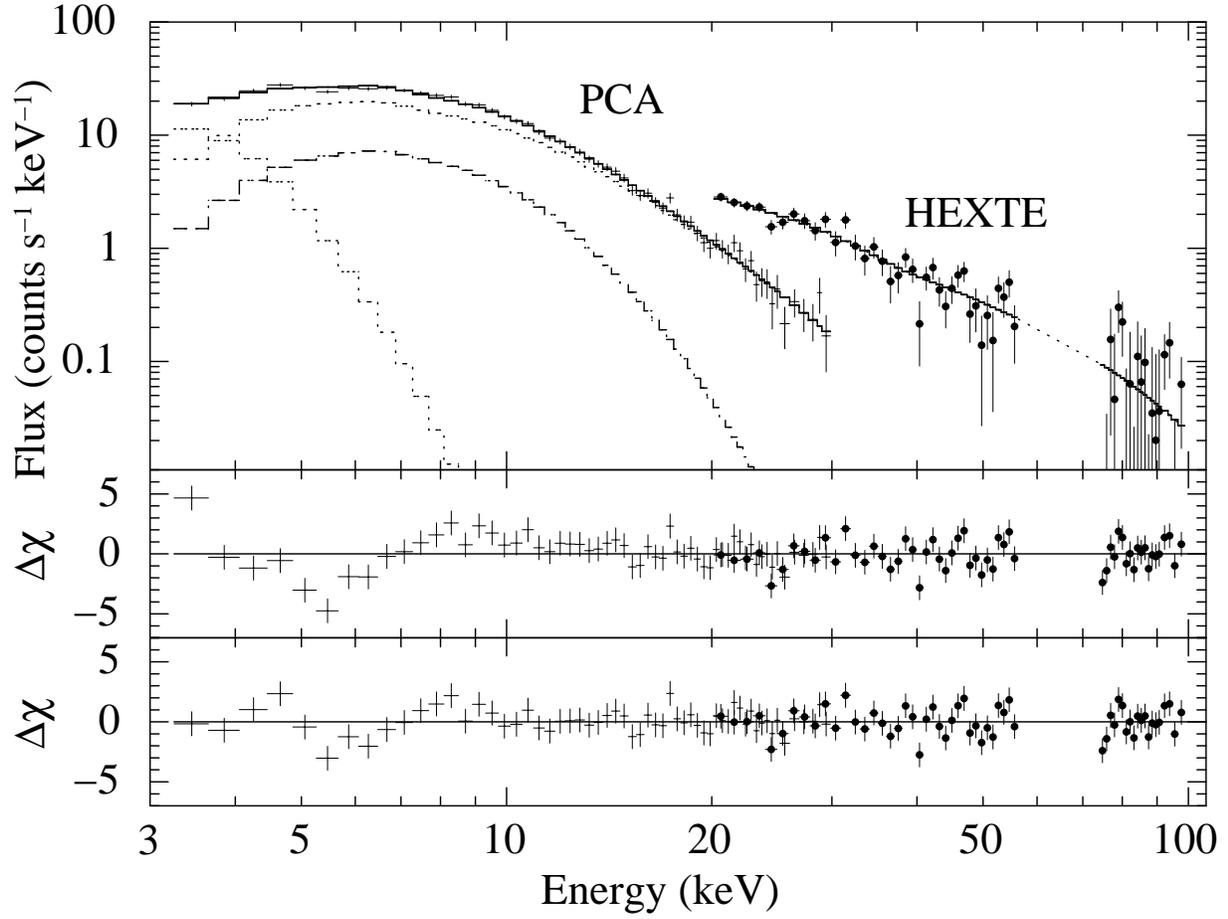}
  \end{center}
  \caption{Spectrum for the dip phase interval fitted with the PL+BB+Fe+$\mathrm{BB_{low}}$ model is shown in top panel. Middle and bottom panels show the residuals for the PL+BB+Fe and PL+BB+Fe+$\mathrm{BB_{low}}$ models, respectively.}\label{fig:spec_bb}
\end{figure}

The top panel of figure \ref{fig:spec_bb} shows the spectrum of the dip phase.
The middle panel presents residuals of the data from the PL+BB+Fe model.
The wavy residuals left in 3$-$6 keV suggest the existence of another component.
We, then, added a low-temperature blackbody component ($\mathrm{BB_{low}}$) with the column density fixed at that of the phase-averaged spectrum of Obs.\,A.
We refer to this new model as PL+BB+Fe+$\mathrm{BB_{low}}$.
The photon index ($\alpha$), the temperature of the BB component ($kT_{\mathrm{BB}}$) and the center energy of Fe component ($E_{\mathrm{Fe}}$) were again fixed at the values for the pulse-phase-averaged spectrum of Obs.\,A.
The PL+BB+Fe+$\mathrm{BB_{low}}$ model effectively reduced the residuals in the 3$-$6 keV energy band as shown in the lowermost panel of figure \ref{fig:spec_bb}.
Then, the $\chi^2_{\mathrm{red}}$ improved from 1.66 to 1.21. 
The best-fit parameters of the two models are presented in table \ref{tab:dip_pars}.
We also tried a model replacing the $\mathrm{BB_{low}}$ component in the PL+BB+Fe+$\mathrm{BB_{low}}$ model with a disk blackbody component (diskBB), then obtained a similar result (table \ref{tab:dip_pars}).

\section{Discussion} 

\subsection{Dip in Pulse Profile}

We found a dip structure in the pulse profile during Obs.\,A, near the peak of the outburst (figure \ref{fig:pulse1}). 
So far, detections of similar dips in the pulse profiles have been reported in two Be/X-ray binaries A 0535+262 and 
RX J0812.4$-$3114, and a low-mass X-ray binary pulsar GX 1+4. 
In the case of A 0535+262, 
a narrow dip structure was detected at the inter-pulse phase in its double-peaked pulse profile during normal outbursts \citep{bildsten1997} and a giant outburst (\cite{caballero2008} and \cite{naik2008}).
RX J0812.4$-$3114 showed a dip in the pulse profile in the 3$-$10 keV band \citep{reig1999a}, whose feature is similar to that of LS V +44 17 in that the pulse profile is sinusoidal and the dip appears in the declining pulse phase.
GX 1+4 has a single-peaked profile with a sharp dip \citep{giles2000}.
The dip in this source was visible in the quiescence as well as in the bright state and the dip duration is shorter in the 7$-$20 keV band than in the 2$-$7 keV band.

The origin of the dip may be interpreted as an eclipse of X-ray emission on neutron star by an accretion flow  (\cite{cem1998} and \cite{gall2001}); X-rays from the surface of the neutron star are absorbed by the accretion stream or the accretion column when it intersects the line of sight as the neutron star rotates.
\citet{cem1998} modeled the emission from a neutron star considering the eclipse and successfully reconstructed the pulse profile of A 0535+262.
The pulse-phase-resolved analysis of A 0535+262 by \citet{naik2008} supports this model, which showed that the column density was largest at the phase of the dip feature.
\citet{gall2001} also showed that the optical depth and the column density were maximum at the dip phase in GX 1+4 and RX J0812.4$-$3114. 
This characteristic of the column density is similar to those of LS V +44 17 (figure  \ref{fig:res_pars}).
Thus, we infer that the dip in the pulse profile of LS V  +44 17 is the accretion column eclipse.

The dip feature only appeared in the pulse profile of Obs.\,A, not in those of Obs.\,B and C. 
The transient nature of the dip may be explained by the change of the accretion rate. 
When the accretion rate is high, the accretion column is broad and cover a large area of the emission region. 
In this situation we can observe the dip in the light curve. 
As the luminosity decreases, the accretion column becomes thinner, then the dip becomes shallower and finally disappears.
Another explanation for the change of the dip is progression of the accretion flow with the orbital motion. 
Since the line of sight varies with the orbital motion, the accretion column may move out of the eclipsing position. 
To discuss more quantitative detail for the sporadic nature of the dip we need to know the ephemeris and the orbital elements of the binary, which we do not have.

\citet{hamada1980} and \citet{nagel1981} proposed
another process to produce a dip structure.
In their models, 
the polarized photons emitted from the polar caps of the neutron star are anisotropically Thomson-scattered
in the magnetized plasma of the accretion column.
The pulse profile is then modified to a complex structure with a dip.
This mechanism can also explain the pulse profile of LS V +44 17.
To distinguish between the anisotropic scattering model and the accretion column eclipse model,
the pulse-phase-resolved X-ray polarimetry would be necessary.

\subsection{Blackbody components in Spectrum}

In all these spectra, there is a 1.8$-$2.2 keV blackbody component (BB).
This component may be emitted from the polar cap of the neutron star because the radius of the BB emission area $R_{\mathrm{BB}} < 1\ \mathrm{km}$ is comparable to that of the accreting polar cap $\sim 0.1R_{\mathrm{NS}}$, where $R_{\mathrm{NS}}$ is the radius of the neutron star ($\sim 10$ km).
In addition, the temperature $kT_{\mathrm{BB}} \sim$ 2 keV, which is higher than the values reported in other accretion pulsars (see table 4 in \cite{2006A&A...455..283L}), suggests that the emission on the neutron star surface reached the local Eddington limit during the high luminosity state.

Another blackbody component $\mathrm{BB_{low}}$, which was found in the dip spectrum, has a relatively low temperature $kT_{\mathrm{BBlow}} \sim 0.5$ keV and a radius of the emission region, $R_{\mathrm{BBlow}} \sim 19\ \mathrm{km}$, larger than that of a neutron star.
According to the physical picture of the soft excess described in \citet{hick2004}, the origin of the soft excess emission may be explained by reprocessing of the hard X-rays by the inner edge of the accretion disk.
However, the radius of the $\mathrm{BB_{low}}$ component is smaller than the typical value of the inner disk radius $R_{\mathrm{in}} \sim 100 R_{\mathrm{NS}}$.
Thus, we suggests the other possible explanation, the emission from the multi-temperature accretion column.
We note that the $\mathrm{BB_{low}}$ component is required only for the dip spectrum. 
It can be explained by such a situation that the $\mathrm{BB_{low}}$ component is always present but dominated by the other components in all the pulse phases except in the dip.

\subsection{Transient Activity of LS V +44 17}

LS V +44 17 has been classified to the class of persistent Be/X-ray binaries because of its longer spin period and the lower X-ray variability than those of transient pulsars \citep{reig1999b}.
The persistent binaries are thought to be a binary systems consisting of a neutron star and a Be star in the relatively wide orbit. 
In these systems, 
a neutron star orbits through the outer region of the circumstellar disk of a Be star
where the matter density is low, thus the mass accretion rate is supposed to be low.
Contrary to this picture,
LS V +44 17 underwent an outburst as we reported.
This change of behavior may be explained by the secular variation of the circumstellar disk of the Be companion.
\cite{reig2005} reported based on the optical/IR monitoring observations of LS V +44 17 that the typical duration of formation/dissipation of the circumstellar disk is more than ten years. 
The outburst in 2010 April may be triggered by mass transfer from the disk which grew large enough over these years to reach the periastron of the neutron star orbit \citep{okazaki2001}.
Axial asymmetry of the circumstellar disk of Be stars may also affect the temporary mass transfer.

Insights on the orbital period can be obtained from the two additional
outbursts from LS V +44 17 on 2010 September 3 (MJD=55442) and 2011 January 31 (MJD=55592) detected by INTEGRAL \citep{krivonos2010} and Swift \citep{tsygankov2011}.
We analyzed the RXTE follow-up observation data of these two outbursts obtained on 2010 September 8-13 and 2011 February 2-7.
The pulse profiles of the RXTE data showed a sinusoidal curve and no significant dip structure in all these epochs.
The intervals between three outbursts (2010 April, September and 2011 January) are both $\sim$ 155 days \citep{tsygankov2011}. If this interval corresponds to the orbital period, this binary system also follows the correlation between the pulse period of neutron star and the orbital period of the Be/X-ray binary, proposed by \citet{corbet1984}.

\section{Conclusion}
The outburst from Be/X-ray binary LS V +44 17 was detected by MAXI on 2010 March 31, which was the first detection of a transient activity in this source. 
The outburst lasted for almost a month, which has been monitored by the MAXI/GSC in the 2$-$20 keV band.
Follow-up observations performed by RXTE revealed the sinusoidal pulse profiles with a period of $\sim$ 205 s.
The pulse profile on 2010 April 6, when the luminosity was near the maximum,
had a sharp dip structure.
The dip was clearer in the lower energy bands.
We infer that an origin of the dip is an eclipse by the accretion column.

In order to verify the physical picture we analyzed the pulse-phase-averaged and resolved spectra of the RXTE data.
The phase-averaged spectra were well represented by a composite model consisting of a cut-off power-law component, a blackbody component with relatively high temperature ($\sim$ 2 keV) and an iron emission line at $\sim$ 6.4 keV.
Considering the size of the emission region, the blackbody component is likely to be emitted from the polar cap on the neutron star.
With the pulse-phase-resolved spectroscopy, we found that the absorption column density at the dip phase was much higher than that at other phases.
This is consistent with the aforementioned picture, 
that the dip seen in the pulse profile of LS V +44 17 is the eclipse by the accretion column.
The spectrum at the dip phase suggests the presence of another soft component
which may be emitted from the multi-temperature accretion column.
\\

This paper is based on use of MAXI data provided by RIKEN and JAXA.
RXTE and Swift data are provided by NASA.
This research was partially supported by the Ministry of Education, Culture, Sports, Science and Technology (MEXT), Grant-in-Aid No.19047001, 20041008, 20244015, 20540237, 21340043, 21740140, 22740120, 23540265 and the Global-COE programs from MEXT ``The Next Generation of Physics, Spun from Universality and Emergence'' and ``Nanoscience and Quantum Physics''.
We thank JAXA and NASA for the operation of MAXI, JEM and ISS.

\bigskip


\begin{table}[p]
  \caption{Summary of RXTE observations.}\label{tab:obs}
  \begin{center}
    \begin{tabular}{cccccccc}
      \hline
      Obs. &Date &  Observation Time & Obs ID & \multicolumn{2}{c}{PCA (3$-$30 keV)\footnotemark[$*$] }& \multicolumn{2}{c}{HEXTE (20$-$100 keV)} \\ \cline{5-6} \cline{7-8} 
      & (2010) & Start-End & (95418-01- ) & Exposure  & Mean Rate\footnotemark[$\dagger$]  & Exposure & Mean Rate\footnotemark[$\dagger$] \\ 
       &  &  (UT)  &    &    (ks)    &   ($\mathrm{counts\  s^{-1}}$) &    (ks)    &   ($\mathrm{counts\  s^{-1}}$)     \\
        \hline
      A & Apr 06 & 07:41 -  09:20& 01-00 & 3.1 & $271.0\pm0.3$ & 2.0 & $40.4\pm0.3$ \\
      B & Apr 12 & 04:47 - 06:29 & 02-00 & 3.3 & $183.8\pm0.2$ & 2.2 & $25.2\pm0.2$ \\
      C & Apr 15 & 05:00 - 07:38 & 02-01 & 3.2 & $91.0\pm0.2$   & 2.2 & $12.2\pm0.2$ \\            
      \hline
    \end{tabular}
  \end{center}
      {\footnotesize
      \footnotemark[$*$] Using PCU2 only. \\
      \footnotemark[$\dagger$] Background is subtracted. 
      }
\end{table}

\begin{table*}[p]
  \caption{pulse-phase-averaged spectral fits results.}\label{tab:spec_pars}
\begin{center}
    \begin{tabular}{lccc}
    \hline
	Parameter & \multicolumn{3}{c}{Value} \\
	                    &  Obs.\,A & Obs.\,B  & Obs.\,C \\	
	\hline
	$N_{\rm{H}}$ $(10^{22}\ \rm{cm}^{-2})$ 		
	& $3.3 \pm0.3$ 	& $3.3\pm0.4$		& $2.0\pm0.6$ \\	
	Power-law index $\alpha$ 				
	& $1.15 \pm 0.04$ 	& $1.22 \pm 0.05$ 	& $1.1\pm0.1$ \\	
	Cut-off energy $E_{\rm{fold}}$ (keV)			
	& $26 \pm 1$ 		& $25 \pm 1$ 		& $20 \pm 2 $ \\	
	Cut-off power-law flux\footnotemark[$*$]
	& $2.05 \pm 0.06$	& $1.40\pm0.05$ 	& $0.62\pm0.05$ \\	
	Blackbody temperature  $kT_{\rm{BB}}$ (keV)	
	& $2.17^{+0.06}_{-0.05}$ 		& $2.07^{+0.07}_{-0.06}$ 		& $1.76\pm0.05$ \\	
	Blackbody radius\footnotemark[$\dagger$] $R_{\rm{BB}}$ (km) 		
	&$0.6 \pm 0.2$		& $0.5\pm0.2$		& $0.6\pm0.3$ \\	
	Iron line energy $E_{\rm{Fe}}$ (keV)			
	&  $6.44^{+0.10}_{-0.09}$ 	& $6.4\pm0.1$				& $6.2^{+0.3}_{-0.2}$ \\	
	Iron line equivalent width $W_{\rm{eq}}$ (eV) 		
	&$43.6^{+5.4}_{-6.1}$		& $35.9^{+6.0}_{-6.9}$		& $17.7^{+8.2}_{-10.2}$ \\		
	Iron line flux \footnotemark[$\ddagger$]		
	&$1.5 \pm 0.3$		& $0.8\pm0.2$		& $ 0.2\pm0.1$ \\	
	Reduced chi-squared\footnotemark[$\S$] $\chi^2_{\rm{red}}$	
	&1.10 (116)		& 1.09 (116)		&  1.01 (116) \\	
	\hline
	\multicolumn{4}{@{}l@{}}{\hbox to 0pt{\parbox{180mm}{\footnotesize
	\par\noindent	
	Uncertainties are 1-sigma confidence limits.
	\par\noindent
      	\footnotemark[$*$] Flux in the energy range 3$-$10 keV, in unit of $10^{-9}\ \rm{ergs}\ \rm{cm}^{-2}\ \rm{s}^{-1}$. 
	\par\noindent
       	\footnotemark[$\dagger$]  Radius for a source distance of 3.3 kpc.
	\par\noindent
	\footnotemark[$\ddagger$] In unit of $10^{-3}\ \rm{photons}\ \rm{cm}^{-2}\ \rm{s}^{-1}$.
	\par\noindent
	\footnotemark[$\S$] Parenthetic values are the degree of freedom.
	}\hss}}
    \end{tabular}
  \end{center}
\end{table*}

\begin{table*}[p]
  \caption{Spectral parameters and $\chi^2_{\mathrm{red}}$ from the dip phase. }\label{tab:dip_pars}
  \begin{center}
    \begin{tabular}{lccc}
      \hline
       Parameter & \multicolumn{3}{c}{Value} \\
       & PL+BB+Fe & PL+BB+Fe+$\mathrm{BB_{low}}$ & PL+BB+Fe+diskBB\\
       \hline
      $N_{\mathrm{H1}}$\ $(10^{22}\ \mathrm{cm}^{-2})$           
      		& $9.7^{+0.4}_{-0.5}$ 
		& $18^{+3}_{-2}$ 
		& $18 \pm 2$ \\
      Cut-off energy $E_{\mathrm{fold}}$\ (keV)                         
      		& $29^{+2}_{-1}$  
		& $30 \pm 2$  
		& $30 \pm 2$ \\
      Cut-off power-law flux\footnotemark[$*$] 
      		& $2.05^{+0.10}_{-0.09}$ 
		& $1.94 \pm 0.10 $ 
		& $1.95 \pm 0.10 $ \\            
      Iron line flux\footnotemark[$\dagger$] 
      		& $<$ 0.15
		& $<$ 0.18
		& $<$ 0.17 \\
      BB radius $R_{\mathrm{BB}}$\ (km)\footnotemark[$\ddagger$]        
      		& $0.4 \pm 0.2$ 
		& $0.7 \pm 0.3$ 
		& $0.7 \pm 0.3$ \\ 
      $N_{\mathrm{H2}}$\ $(10^{22}\ \mathrm{cm}^{-2})$           
      		& $-$ 
		& $3.3$ (fixed) 
		& $3.3$ (fixed) \\            
      $\rm{BB_{low}}$ temperature $kT_{\mathrm{BBlow}}$\ (keV)
      		& $-$
		& $0.48^{+0.06}_{-0.07}$
		& $-$ \\
      $\rm{BB_{low}}$ radius $R_{\mathrm{BBlow}}$\ (km)\footnotemark[$\ddagger$]          
      		& $-$ 
		& $19^{+25}_{-13}$
		& $-$\\ 
      diskBB temperature $kT_{\mathrm{in}}$\ (keV) 
      		& $-$
		& $-$
		& $0.55\pm0.09$ \\
     diskBB radius $R_{\mathrm{in}} \sqrt{\cos \theta}$\ (km)\footnotemark[$\ddagger$]          
      		& $-$ 
		& $-$ 
		& $16^{+26}_{-12}$ \\ 
      Reduced chi-squared\footnotemark[$\S$] $\chi^2_{\mathrm{red}}$ 
      		& 1.66  (111)
		& 1.21  (109)
		& 1.21  (109) \\                                             
      \hline
      	\multicolumn{4}{@{}l@{}}{\hbox to 0pt{\parbox{180mm}{\footnotesize
	$N_{\mathrm{H1}} =$ hydrogen column density affecting PL.  $N_{\mathrm{H2}} =$ hydrogen column density affecting $\rm{BB_{low}}$ or diskBB.
	\par\noindent	
	Uncertainties are 1-sigma confidence limits.
	\par\noindent
  \footnotemark[$*$] Flux in the energy range 3$-$10 keV, in unit of $10^{-9}\ \rm{ergs}\ \rm{cm}^{-2}\ \rm{s}^{-1}$.
	\par\noindent  
  \footnotemark[$\dagger$] Flux of the iron line in unit of $10^{-3}\ \mathrm{photons\ cm}^{-2}$ $\mathrm{s}^{-1}$.
  	\par\noindent
  \footnotemark[$\ddagger$] Radius for a source distance of 3.3 kpc. $\theta$ is the angle of accretion disk.
  	\par\noindent
  \footnotemark[$\S$] Parenthetic values are the degree of freedom.
  }\hss}}
    \end{tabular}
  \end{center}
\end{table*}

\end{document}